\documentstyle[12pt,fleqn,cite]{article}
\newcommand{\sect}[1]{\setcounter{equation}{0}\section{#1}}
\newcommand{\subsect}[1]{\subsection{#1}}
\newcommand{\subsubsect}[1]{\subsubsection{#1}}
\renewcommand{\theequation}{\thesection.\arabic{equation}}
\def\sxn#1{\sect{#1}}
\def\subsxn#1{\subsect{#1}}
\def\subsubsxn#1{\subsubsect{#1}}
\font\mybb=msbm10 at 12pt
\def\bb#1{\hbox{\mybb#1}}
\def\K {\bb{K}}
\def\S {\bb{S}}
\def\Z {\bb{Z}}
\def\T {\bb{T}}
\def\be{\begin{equation}} 
\def\ee{\end{equation}}
\def\bea{\begin{eqnarray}} 
\def\eea{\end{eqnarray}}
\def\pa{\partial} 
\def\dmu{\partial_\mu} 
\def\rt{\rightarrow} 
\begin{document} 
\title{ \vspace{-2cm}\flushright{\small IP/BBSR/96-29 \\ hep-th/9604164} \\
\vspace{2cm} Compactifications of M-Theory  to Two Dimensions}
\author{Alok Kumar\thanks{kumar@iopb.ernet.in} 
and Koushik Ray\thanks{koushik@iopb.ernet.in} \\
 Institute of Physics, \\Bhubaneswar 751 005, INDIA} 
\date{\today}
\maketitle
\begin{abstract}
\noindent Compactifications of M-theory to two dimensional space-time
on ${(K3\times \T^5)}/ \Z_2$ and 
${(K3\times K3\times \S^1)}/ \Z_2$ orientifolds are presented. 
These orientifolds provide examples of anomaly free
chiral supergravity models in two dimensions
with (8, 0) and (4, 0) supersymmetries. 
Anomaly free spectra at the enhanced symmetry points are also 
obtained. The results confirm the twisted sector contribution to
the spectrum in the case of $\T^9/ \Z_2$ discussed earlier.  
\end{abstract}

\sxn{Introduction}
Various compactifications of M-Theory\cite{mth,mth2,maha,horav,mukhi,wit2,
mtdual,davis,sen,ray,aharony,acharya,
tsen,verlinde,ferrara,aldabe} have been discussed
recently in order to understand the duality symmetries
of string theory in a unified framework. In particular a great
deal of attention has been paid to the ten\cite{horav} and six
dimensional\cite{mukhi,wit2,sen,ray,verlinde} 
cases due to the nontrivial anomaly cancellation
requirements in these dimensions. Since the full content of
M-Theory is not known, the anomaly cancellation\cite{alvarez,anomaly} 
provides useful
informations about its consistent compactifications. In
particular this can be used to 
predict the contribution of the twisted sectors
of these theories. For example, in the case of M-Theory
on the orientifold $\S^1/ \Z_2$\cite{horav}, one gets the
$E_8$ gauge multiplets from each of the two twisted sectors of the
theory which gives the $E_8\times E_8$   
heterotic string in ten dimensions.  
In the six dimensional case, one encounters a number of
possibilities. In this case chiral supergravity theories can be
constructed either with $N=1$ or $N=2$ supersymmetries. For
the $N=2$ case the anomaly cancels for one gravity and
21 tensor multiplets. An explicit
construction of this type can be obtained from $M$-Theory when
it is compactified on a $\T^5/ \Z_2$ orientifold. In the $N=1$ case,
there are several ways that this restriction is satisfied. One
such possibility is for one gravity, 9 tensor,
$m$ $U(1)$ vectors and $(m+12)$ hyper-multiplets\cite{anomaly}. 
These are also anomaly free when the symmetry enhances at
special points of the moduli space in these theories 
in certain directions\cite{sen}. An explicit
example for a model of this type for $m=0$ is obtained by type IIB
string compactification on an orientifold $K3/ \Z_2$\cite{dabol,gimon}. 
$M$-Theory
compactification on $K3\times \S^1/ \Z_2$\cite{sen,ray} 
and their further
projections give many such models with different number of vector
multiplets. 

Another case, where the anomaly cancellation puts nontrivial 
restrictions on the construction of consistent theories is 
two-dimensions. The anomaly contributions of the 
spin-$1\over 2$ and spin-$3\over 2$ Weyl spinors  
in two dimensions are respectively\cite{alvarez}, 
\[I_{1\over 2} = -{1\over 24} p, \qquad I_{3\over 2} =  
{23\over 24} p,  \]
where $p$ is the anomaly polynomial.
In addition, a chiral right-moving scalar also contributes to anomaly
in two dimensions 
which is equal to that of the positive chirality 
spin-$1\over 2$ Weyl field. Examples of 
anomaly cancellations appear in the context of the first quantized
string constructions.   
For known string theories with $(p, q)$ supersymmetries, the anomaly trivially 
cancels when $p=q$. For $p\neq q$, it is well 
known that the gravitational anomaly on the string worldsheet 
cancels  for the field content
of the heterotic string. In this case, the 
anomaly cancellation gives the difference between the numbers of 
left and right-moving fermionic coordinates (=22)  
in the heterotic string theory in its fermionic formulation. 

Anomaly cancellation in two dimensions in the context of $M$-Theory
compactification were discussed in \cite{mukhi} where 
construction of these theories on $\T^9/ \Z_2$
orientifolds have been discussed. 
A new feature of the $\T^n/ \Z_2$ orientifold was noticed in
\cite{mukhi}, where it was argued that the fixed points of the
discrete group are not uniformly distributed along the sides of
the torus, implying a compactification on degenerate torii. It was
conjectured that the  
only exception was the two dimensional case where all
the 512 Majorana-Weyl
({\sf mw}) fermions necessary for anomaly cancellation 
arise from 512 twisted
sectors of $\T^9/\Z_2$\cite{mukhi,wit2}. This provides an example of (16, 0) 
supergravity model in two dimensions.

In view of the significance of $M$-Theory in understanding
the duality symmetries and other aspects of string theories,  it
is important to analyze other compactifications of this theory. 
In this paper, we discuss other orientifold
compactifications to two dimensions, namely $(K3\times \T^5)/ \Z_2$
and $(K3\times K3\times \S^1) / \Z_2$. By defining the  
$ \Z_2$ symmetry on the eleven-dimensional fields as well as on the 
compactified coordinates, including the harmonic forms on $K3$,
we explicitly compute the
fermionic and bosonic zero modes for these cases. Then by
combining these with the twisted sector states, namely 256 and
128 {\sf mw} fermions in the two cases, the anomaly
cancellation condition is satisfied. 
Our results confirm the contribution of the twisted sector states 
mentioned above for $\T^9/ \Z_2$ in
\cite{mukhi,wit2}. We also study the enhanced symmetry points in these
cases using the equivalence of $M$-Theory on $K3\times \S^1$ 
with the heterotic string on $\T^4$. 
We find the particle spectrum at
these points and once again show the anomaly cancellation.

\sxn{$\T^9/ \Z_2$ orientifold}

We now begin by reviewing the case of $M$-Theory on 
$\T^9/ \Z_2$ and  
apply the techniques to the $(K3\times \T^5)/ \Z_2$ and 
$(K3\times K3\times \S^1)/ \Z_2$ orientifolds subsequently.
We shall find both the bosonic and the fermionic spectrum for 
the $\T^9/ \Z_2$ orientifold. 
They coincide with the one mentioned in \cite{mukhi,wit2} on 
the basis of supersymmetry multiplet structure.  
First, let us  present the  spectrum for  compactification
on $\T^9$. Then by defining the $ \Z_2$ projection we obtain the 
spectrum for $\T^9/ \Z_2$. Since the only bosonic fields that enter into
the counting of the physical degrees of freedom are the graviton and
the scalars, we restrict to these fields only. 
For toroidal 
compactification, the bosonic spectrum in two dimensions consists
of $G_{\mu \nu}$ and $G_{a b}$ components of the the 
eleven-dimensional field $G_{M N}$. These give one graviton and 
45 scalars in two dimensions. 
Only two dimensional fields originating from the 
eleven dimensional fields 
$A_{M N P}$ in two dimensions are the 84 scalars $A_{a b c}$. Together  
we have 1 graviton and 129 scalars.
Since a graviton in two dimensions has $(-1)$ physical degree of freedom
due to gauge fixings\cite{mukhi},
they have to be taken into account in counting the number of scalars. 
Effectively 
only bosonic degrees of freedom are that of 128 scalars.

The fermionic degrees of freedom can be counted by decomposing the
eleven dimensional Majorana spin-$3\over 2$ spinor into a set of 
spin-$1\over 2$ and spin-$3\over 2$ spinors in two dimensions. 
We find it more convenient to decompose the eleven dimensional 
Majorana spinor first in terms of the ten dimensional 
{\sf mw} spin-$3\over 2$ fields $\Psi^{\pm}_{\bar{\mu}}$ and 
and spin-$1\over 2$ fields 
$\Psi^{\pm}_{10}$. The two dimensional field content is then 
found by decomposing these as a product of the 2d space-time spinor
and an internal eight-dimensional spinor. 
$\Psi^{+}_{\bar{\mu}}$ can be decomposed into a set of
two dimensional spin-$3\over 2$ fields $\psi_{\mu}^+ \chi^+_8$,
$\psi_{\mu}^- \chi^-_8$  and spin-$1\over 2$ fields $\psi^+ \chi_{8 i}^+$,
$\psi^- \chi_{8 i}^-$, $\mu = 0, 1$ and $i= 1..8$, where $\chi_8^{\pm}$
($\chi_{8 i}^{\pm}$) are the spin-$1\over 2$ ($3\over 2$) modes on $\T^8$.
As a result  $\Psi^{+}_{\bar{\mu}}$ yields 
8 gravitinos ($\psi_{\mu}^+$) of positive chirality 
and 8 gravitinos ($\psi_{\mu}^-$) of negative chirality in two dimensions. 
We also have 64 {\sf mw} spinors ($\psi^+$) of positive 
chirality and 64 of negative chirality in two dimensions. 
Similarly $\Psi^-_{\bar{\mu}}$
contributes 8 two-dimensional gravitinos of positive and 
8 of negative chirality and gives 64 {\sf mw}
spinors of positive and 64 of negative chirality. The ten dimensional
{\sf mw} spinor $\Psi^+_{10}$ can be decomposed as a product 
$\psi^+ \chi_8^+$ and $\psi^- \chi_8^-$. 
It gives 8 {\sf mw} spinors of
positive and 8 of negative chirality. The ten dimensional 
{\sf mw} spinor $\Psi^-_{10}$ can also be decomposed into 8  positive and
8 negative chirality two dimensional {\sf mw} spinors. To count the 
physical degrees of freedom, we note that each of the gravitinos
in two dimensions has a $(-1)$ degrees of freedom\cite{mukhi} and 
cancels the physical degree of freedom of a {\sf mw} 
spinor of opposite chirality. As a result 
the physical degrees of freedom in two dimensions consists of
128 {\sf mw} spinors of positive chirality and 128 of negative chirality.
Combining them with the 128 scalars they form 8 scalar multiplets of
(16, 16) supersymmetry in two dimensions. This is the maximal 
supersymmetry that can be obtained in two dimensions through
compactifications of known higher dimensional theories. 
An infinite dimensional symmetry structure for these theories 
has been demonstrated and
their integrability has been argued\cite{nicolai}. 
The coset space structure 
for this theory is known to be ${E_{8(8)}/ SO(16)}$ which is 
parameterized by 128 scalars found above.  

We next discuss the model which is obtained by an orientifolding
of the one discussed above. The orientifolding is defined as
a $ \Z_2$ projection which acts on the eleven-dimensional 
3-form field by a change of sign, {\it viz.}
 $A_{M N P}\rightarrow - A_{M N P}$. 
At the same time the  11-dimensional gravitino 
transforms as $\psi_M \rightarrow 
\gamma^{11}\psi_M$. 
Here $\gamma^{11} $ is the analogue of the four-dimensional 
$\gamma^5$ and is defined as the product of ten-dimensional 
gamma matrices, $\gamma^{11} \equiv \gamma^0 \cdots \gamma^9$.
For the ten dimensional fields it implies,
$\Psi^{\pm}_{\bar{\mu}} \rightarrow \pm \Psi^{\pm}_{\bar{\mu}}$,
$\Psi_{10}^{\pm} \rightarrow \mp\Psi_{10}^{\pm}$.
The $ \Z_2$ symmetry also acts on the internal coordinates as
$(x^2,...,x^{10}) \rightarrow - (x^2,...,x^{10})$. The result
of this operation on the fermionic coordinates can be 
obtained in various ways. A simple way to see their effect on the
fermions is to notice that, in order to keep
the worldsheet supercurrent single valued,
the worldsheet fermionic 
coordinates are also odd under the above symmetry.
Then by writing down the vertex operators for the spinors in 
terms of the bosonized fields $y_1,..,y_4$ one finds, 
$\chi_8^{\pm} \rightarrow \chi_8^{\pm}$ and 
$\chi_{8 i}^{\pm} \rightarrow {\mp}\chi_{8i}^{\pm}$. 

States in the untwisted sector consists of those left invariant under 
this $ \Z_2.$ All the 
129 scalar fields as well as the graviton survive the $ \Z_2$ projection.
Among the fermions, the 
set of fields invariant under $ \Z_2$ are 16 {\sf mw}
spin-$3\over 2$ states with positive chirality and 144 {\sf mw} 
spin-$1\over 2$ states of negative chirality. 16 of these
spin-$1\over 2$ states 
cancel the negative physical degrees of gravitinos.
The presence of 
16 gravitinos of positive chirality implies that the result of
orientifolding of $\T^9$ compactification gives a 
(16, 0) local supersymmetric model in two dimensions. As expected,
the number of supersymmetries reduces by a factor of half with 
respect to the compactification $\T^9$. 

The total anomaly of the 144 negative chirality 
 spin-$1 \over 2$ and 16 positive chirality spin-$3 \over 2$
is given by 
\[ I = - 144 I_{1\over 2} + 16 I_{3\over 2}
 = {512\over 24} p. \]
The anomaly is cancelled by 
512 {\sf mw} spin-$1\over 2$ fields of positive chirality. 
On the basis of this fact, it
was argued  in \cite{mukhi,wit2} that the twisted sectors
contribute 512 {\sf mw} spinors, each one contributing
one such fermion. It has also been pointed out\cite{wit2} that 
these fermions are singlets under the $(16, 0)$ supersymmetry 
mentioned above. As a result, the contribution of each of the 
twisted sector still comes as a supersymmetry multiplet  
and the 512 extra scalars mentioned in \cite{mukhi} 
are not needed for maintaining the supersymmetry structure.

It is also known\cite{salam} that 
a truncation of the $d=3$, $N=16$ supergravity leads to a
(16,0) supersymmetric model in two dimensions with the 
coset structure of the moduli fields being $E_{8 (8)}/SO(16)$.
This result coincides with the number of scalars found above. 
\sxn{Orientifold of $K3\times \T^5$}
We now describe the  compactification of $M$-theory on $K3\times \T^5$ 
and obtain its orientifold by a $ \Z_2$ projection and
by adding the twisted sector states. As in the case of $\T^9$,
we shall ignore the one and higher-form fields in the bosonic spectrum.
The necessary informations
needed to obtain the bosonic sector of the compactified theory
are the moduli space dimension of the Einstein metric on $K3$, 
namely 58. In addition one also needs the information that the
only nonzero betti numbers for $K3$ are $b_0 = b_4 = 1$ and 
$b_2 = 22$. Out of the 22 two-forms on $K3$, there are 3 self-dual
and 19 anti-self-dual 2-forms. In addition we also use the 
fact that $K3$ does not have any continuous isometries. 
From the eleven-dimensional 
graviton $G_{MN}$ one now obtains, in two dimensions, 
one graviton $G_{\mu \nu}$ and  58 scalars $G_{m n}$, where $m$ is 
symbolically representing $K3$. In addition, one also
has 15 scalars $G_{a b}$, where $a, b$ etc. are the indices 
on $\T^5.$ By compactifying the 
eleven-dimensional three form $A_{MNP}$, one obtains, in 
two dimensions, 110 scalars $A_{m n a}$ and 10 
scalars $A_{a b c}$. Combining these results we have a graviton and 
193 scalars. As stated earlier, in two dimensions, one of the
scalars compensates for the negative degree of freedom of  the 
graviton and one is finally left with the 192 physical scalar degrees 
of freedom.

We will now find the fermionic degrees of freedom for this case.
For this computation one needs the the index of the spin-$1\over 2$ 
Dirac operator and 
spin-$3\over 2$ Rarita-Schwinger operator on $K3$\cite{walton}. The 
spin-$1\over 2$ Dirac index on $K3$ is given by 
\[I^-_{1\over 2} - I^+_{1\over 2} = 2,\]
and the spin-$3\over 2$ index is given as  
\[I^-_{3\over 2} - I^+_{3\over 2} = -40.\]

The two dimensional fermionic spectrum is now obtained in 
the same manner as in the last section by decomposing the ten dimensional
{\sf mw} fermions $\Psi^{\pm}_{\bar{\mu}}$ and $\Psi_{10}^{\pm}$ into a 
product of a two dimensional spinor, a spinor on $K3$ and another
one on $\T^4$. We denote  
the spin-$3\over 2$ ($1\over 2$) {\sf mw} fields in the two dimensional 
space-time by $\psi_{\mu}^{\pm}$ ($\psi^{\pm}$);
on $K3$ by $\chi^{\pm}_{k m}$ ($\chi^{\pm}_k$) and 
on $\T^4$ by $\chi_{4 i}^{\pm}$ ($\chi^{\pm}$). 
Then from $\Psi_{\bar{\mu}}^{\pm}$ we obtain four
gravitinos ($\psi_{\mu}^- \chi_K^- \chi_4^+$) of negative chirality
and four gravitinos ($\psi_{\mu}^+ \chi_{k}^- \chi_4^-$) of positive
chirality using the index of the spin-$1\over 2$ fields on $K3$. 
One finds 80 spinors of positive chirality and
80 of negative chirality as $\psi^+ \chi_{k m}^+ \chi_4^+$ 
and $\psi^- \chi_{k m}^+ \chi_4^-$ respectively
from contributions of the spin-$3\over 2$ index on $K3$.
We also get 16 spinors of negative and 16 of positive chirality 
as $\psi^- \chi_k^- \chi_{4 i}^+$ and 
$\psi^+ \chi_k^- \chi_{4 i}^-$.  Similarly the other 
{\sf mw} spin-$3\over 2$ field in ten dimensions, namely 
$\Psi_{\bar{\mu}}^-$ 
gives 4 positive and 4 negative chirality {\sf mw} gravitinos
in two dimensions. Once again we get 80 positive
and 80 negative chirality spin-$1\over 2$ fields as contribution of
the spin-$3\over 2$ index on $K3$ and 
the index of the spin-$1\over 2$
fields on $K3$ implies that there are 16 spinors of both chiralities.
Using the spin-$1\over 2$ index on $K3$, one can also decompose the 
ten dimensional spinors $\Psi_{10}^+$ and $\Psi_{10}^-$. 
In each case one obtains
4 positive and 4 negative chirality spinors in two dimensions. 

Thus we obtain, in two dimensions, a theory
with eight positive and eight negative chirality spin-$3\over 2$ 
fields. Therefore we have an (8, 8) supersymmetric model in two
dimensions. This model can also be obtained by a 
toroidal compactification of the heterotic string theory to two 
dimensions. This also follows from the fact that the $M$-Theory
compactified to six dimensions on $K3\times \S^1$ is the 
dual description of the heterotic string theory compactified on 
$\T^4$. In fact, the number of scalars in two dimensions, 
{\it viz.} 192, is the same as that for the heterotic string theory
on $\T^8$. The moduli space in this case
is parameterized by $O(8, 24)/ O(8)\times O(24)$. Among 
spinor degrees of freedom,  
8 positive and 8 negative chirality ones are used to cancel the
negative degrees of the gravitino modes. One is finally left
with 192 positive and 192 negative chirality spin-$1\over 2$ fields 
and all of them combine into 24 multiplets of the $(8, 8)$
supersymmetry algebra in two dimensions.  

After discussing the compactification on $K3\times \T^5$, we now obtain
its orientifold through a $ \Z_2$ projection and by adding 
appropriate twisted sectors. The $ \Z_2$ action on the ten dimensional
field was already defined in section-2, namely 
$A_{M N P} \rightarrow - A_{M N P}$ and 
$\psi_M \rightarrow \gamma^{11}\psi_M$. It acts on 
$\T^5$ as $x^a \equiv (x^6,\cdots ,x^{10})\rightarrow - (x^6,\cdots
 ,x^{10})$.
This has the effect of transforming the 
spinors on $\T^4$, namely the internal part of the ten-dimensional 
{\sf mw} spinor discussed earlier, as
$\chi_4^{\pm}\rightarrow {\mp} \chi_4^{\pm}$
and  $\chi_{4 i}^{\pm} \rightarrow \pm \chi_{4 i}^{\pm}$.  

The action of $ \Z_2$ on the bosonic modes of $K3$ is the same as 
the one discussed in \cite{chaudh,sen,ray}. Under this symmetry, 
34 of the 58 $K3$ moduli fields are even and 24 are odd. 
It also leaves 14 two-forms, including all the 
three self-dual two-forms invariant. Remaining 8 two-forms
are odd whereas zero and four-forms are even. 
To describe the $ \Z_2$ operation on
the fermionic zero modes, one exploits their relations
to the harmonic forms on the compact space. 
For example, the number of spin-$1\over 2$ 
modes equals that of the (0, p) forms and the number of 
spin-$3\over 2$ modes equals twice the number of  
 the (1, 1) forms\cite{walton}. As a result, all the 
spin-$1\over 2$ modes on $K3$ coming from the (0,0)- and (0,2)-
forms are invariant under the $ \Z_2$ symmetry.
On the other hand, only 24 of the 40 spin-$1\over 2$ modes 
coming from the 12 invariant (1,1)-forms are 
even under $ \Z_2$ with 16 being odd. Using these informations,
one can compute the number of 
bosonic and fermionic modes which survive the $ \Z_2$ projection. 
One finds that 
the graviton $G_{\mu \nu}$ is invariant under this symmetry. 
Similarly 34 of the $K3$ moduli ($G_{m n}$)
and 15 scalar modes on $\T^5$, namely
$G_{a b}$ and $G_{10, 10}$, are invariant. 
The components $A_{m n a}$ and $A_{a b c}$ contribute
70 and 10 scalar fields in two dimensions respectively. 
By cancelling one of the scalars with the negative degree of 
freedom of the graviton we are 
left with 128 scalars in two dimensions. 

The $ \Z_2$ operation on the spinors leaves the following
set of two-dimensional fields invariant. From the {\sf mw} spinor 
component $\Psi_{\bar{\mu}}^+$ we get the following set of 
invariant fields, (i) 4 positive chirality spin-$3\over 2$ components 
($\psi_{\mu}^+ \chi_k^- \chi_4^-$) coming, {\em e.g.}
from the 2 (0, p) forms on
 $K3$ and 2 surviving Killing spinors on $\T^5$, 
(ii) 16 spin-$1\over 2$ components 
($\psi^- \chi_k^- \chi_{4 i}^+$), (iii) 32 spin-$1\over 2$ positive
chirality components ($\psi^+ \chi_{k m}^+ \chi_4^+$),
and (iv) 48 negative chirality spin-$1\over 2$ components
($\psi^- \chi_{k m}^+ \chi_4^-$).
Similarly from $\Psi_{\bar{\mu}}^-$
we get (i) 4 positive 
chirality spin-$3\over 2$ fields ($\psi_{\mu}^+ \chi_k^- \chi_4^+$),
(ii) 32 positive chirality spin-$1\over 2$ fields 
($\psi^+ \chi_{k m}^+ \chi_4^-$), (iii) 16 negative chirality 
spin-$1\over 2$ fields ($\psi^- \chi_k^- \chi_{4 i}^-$) and 
(iv) 48 negative chirality
spin-$1\over 2$ fields ($\psi^- \chi_{k m}^+ \chi_4^+$). The 
ten-dimensional fields $\Psi_{10}^{\pm}$ have the surviving 
components in two dimensions as $\psi^- \chi_k^- \chi_4^+$
and $\psi^- \chi_k^- \chi_4^-$ respectively. 
As a result, we have an $(8, 0)$ chiral supergravity
model in two dimensions coupled to a set of matter multiplets,
{\it i.e.}
\begin{itemize}
\item{$ (1 g_{\mu \nu}, 1 \phi, 8 \psi^+_\mu, 8 \psi^-$)}
\item{ $(128 \phi_L, 128\psi^-)$}
\item{ $(128 \phi_R, 64\psi^+)$}.
\end{itemize} 
The presence of 64 positive chirality fermions is consistent with
the (8, 0) supersymmetry as it 
has been pointed out earlier, in another context\cite{nishino}, 
that the chiral fermions on the 
worldsheet can be coupled to the $(8, 0)$ supergravity theory.  
The situation is similar to the case of the heterotic string 
theory where there are extra worldsheet {\sf mw} fermions of a 
particular chirality. 
 
We have a surplus of 64 {\sf mw}
fermions with negative chirality. A computation of anomaly contributions
of different fields implies that it can be cancelled by
256 positive-chirality fermions from the twisted
sectors of the theory. Since $K3\times \T^5$ have 
exactly 256 twisted sectors and the contribution of the 
individual twisted
sector remains unchanged with respect to the compactification on 
$\T^9/ \Z_2$ as in six dimensions\cite{sen}, we get precise anomaly 
cancellation in this model with $(8, 0)$ supersymmetry. 
Once again the twisted sector fermions are invariant under this
supersymmetry. 
Next we analyze the case of enhanced symmetry and show that the 
theory is anomaly free in this case as well.

To show that the anomaly cancels at the enhanced
symmetry point also, we follow the approach 
of ref.\cite{sen} in the six dimensional case. 
Symmetry enhancement takes place at points in the moduli space where
there are extra massless multiplets. It has been argued in 
\cite{sen} that such multiplets, considered below, 
do not receive any quantum correction to their masses. 
The enhanced symmetry
in the $M$-Theory compactification can be understood in terms of 
the strong-weak duality between the type IIA string theory
on $K3$ and the heterotic string theory on $\T^4$.  
All the states of the $M$-Theory on 
$K3\times \T^5$ are identified in terms of the heterotic string 
spectrum on $\T^8$. $ \Z_2$ operations are also identified 
in the two cases. Then the $ \Z_2$-projection in 
the heterotic string gives the enhanced symmetry for the $M$-Theory case. 
We also exploit the fact that the twisted sector spectrum does not 
change at the enhanced symmetry point\cite{sen}.  

To write down the states in the heterotic string theory,
we denote the space-time coordinates in the 
left-moving sector as ($X^{L \mu}$, $\psi^{L \mu}$) and the internal 
eight-dimensional  
coordinates as ($X^{L i}$, $\psi^{L i}$). The 
right-moving coordinates are denoted 
as $X^{R \mu}$, $X^{R i}$ and $X^{R I}$, where $X^{R I}$ are
the sixteen extra internal coordinates in the right-moving 
sector. The momentum 
lattice in the internal eight-dimensional space is denoted by a sixteen 
dimensional momentum vector $\vec{k}$. Similarly the $E_8\times E_8$
root vectors are denoted by $\vec{k_1}$ and 
$\vec{k_2}$ respectively. Then the states of the $E_8\times E_8$ 
heterotic string theory, when it is compactified to two dimensions, are
given as:  (i) $\psi^{L \mu}_{-{1\over 2}}|{0}>\otimes 
X^{R \nu}_{-1}|\vec{0}, \vec{0}, \vec{0}>$,
where the vacuum in the left-moving sector is denoted by 
$|0>$ and in the right-moving sector by 
$|\vec{0}, \vec{0}, \vec{0}>$, with the entries denoting the 
momenta along the internal
16-dimensional lattice and two $E_8$ directions  respectively.    
(ii) Scalar states:
(a) $\psi^{L i}_{-{1\over 2}}|{0}>\otimes 
X^{R j}_{-{1}}|\vec{0}, \vec{0}, \vec{0}>$, 
(b) $\psi^{L i}_{-{1\over 2}}|{0}>\otimes 
X^{R I}_{-1}|\vec{0}, \vec{0}, \vec{0}>$, 
(c) $\psi^{L i}_{-{1\over 2}}|0>\otimes 
|{\vec{0}, \vec{\alpha}, \vec{0}}>$, and  
(d) $\psi^{L i}_{-{1\over 2}}|0>
\otimes |{\vec{0}, \vec{0}, \vec{\alpha}}>$, where $\vec{\alpha}$ 
is a root vector of $E_8$. As a result we have 8 scalars in 
the adjoint representation of $E_8\times E_8$. In addition we also 
have 64 scalars which are singlets of $E_8\times E_8$. 
At a generic point in the moduli space, 
we have only 192 scalars, namely (ii.a)
and (ii.b). Together they give the  
coset $O(8, 24)/O(8)\times O(24)$ of the moduli fields. 
Note that the state (i) includes the dilaton. As a result we did not 
need to cancel any negative degree of freedom in this case. Same is
true for the counting of fermions as well. 

Fermions at the enhanced symmetry points can also be counted in 
exactly the same manner. The fermion states of the heterotic string 
theory in two dimensions are:
${(i)}^{\prime}$ $|{\bar{\psi}}> \otimes 
X^{\nu}_{-1}|{\vec{0}, \vec{0}, \vec{0}}>$,
where $|{\bar{\psi}}>$ is the left-moving spinor vacuum in the 
NSR formulation of the heterotic string theory. These are the 16  
two-dimensional gravitino states, 8 with positive and 8 with negative
chirality. ${(ii)}^{\prime}$ 64 spinor states of positive and 64 of 
negative chirality:
$|{\bar{\psi}}> \otimes X^i_{-1}|{\vec{0}, \vec{0}, \vec{0}}>$.
${(iii)}^{\prime}$ 8 positive and 8 negative chirality spinors 
in the adjoint representation of $E_8\times E_8$:
(a)$|{\bar{\psi}}> \otimes X^{R I}_{-1}|{\vec{0}, \vec{0}, \vec{0}}>$,
(b)$|{\bar{\psi}}> \otimes |{\vec{0}, \vec{\alpha}, \vec{0}}>$, and
(c)$|{\bar{\psi}}> \otimes |{\vec{0}, \vec{0}, \vec{\alpha}}>$.

The $ \Z_2$ symmetry in the heterotic string theory
can be defined as a product of
two $ \Z_2$'s, namely $\tau \otimes \sigma$, where $\tau$ is the 
inversion of the $(8, 24)$ lattice and $\sigma$ exchanges
the two $E_8$ lattice vectors.
The momentum vectors defining the roots of
the two $E_8$'s are interchanged under the $ \Z_2$ as 
$(\vec{k}, \vec{k_1}, \vec{k_2}) \rightarrow ( - \vec{k}, -\vec{k_2},
-\vec{k_1})$. 
16-dimensional vector defining the internal coordinates of $\T^8$ changes
sign under this operation, i.e., $x^i \rightarrow -x^i$ for $i=1,..,8$. 
Its operation
on fermions can be found by decomposing them into a two-dimensional and 
an eight-dimensional part. They are given by the decomposition of 
$|\bar{\psi}^+>$ as  $|\psi^+ \chi^+_8>$ and $|\psi^-\chi^-_8>$, with 
$|\chi_8^+>$ ($|\chi^-_8>$) being even (odd) under $ \Z_2$.

Then the $ \Z_2$-invariant bosonic states are the (a) graviton, 
(b) 64 scalars in (ii.a) above which are also singlets of the 
surviving nonabelian group $E_8$. In addition we have the states
in the adjoint representation of the surviving $E_8$. 
They are the linear combinations of the states in (ii.b), (ii.c) and 
(ii.d) above and can be written as 
(c) $\psi^i_{-{1\over 2}}|0>\otimes 
(|{\vec{0}, \vec{\alpha}, \vec{0}}>
- |{\vec{0}, \vec{0}, \vec{-\alpha}}>)$, and 
(d) $\psi^i_{-{1\over 2}}|\vec{0}>
\otimes X^{+ I}_{-1}|{\vec{0}, \vec{\alpha}, \vec{0}}>$, 
where $X^{\pm R I}$ 
is defined as $(X^{(1)R I}_{-1} \pm X^{(2)R I}_{-1})$, with 
$X^{(1)R I}$ and $X^{(2)R I}$ denoting the coordinates in the 
Cartan subalgebra of the first and
second $E_8$ respectively. The $ \Z_2$ invariant fermionic states are 
the eight positive chirality gravitinos:
(a) $|{\psi^+ \chi_8^+}>\otimes X^{\mu}_{-1}|{\vec{0}, \vec{0}, \vec{0}}>$,
(b) 64 spinors of negative chirality which are singlets of 
$E_8\times E_8$:
$|{\psi^- \chi^-_8}>\otimes X^i_{-1}|{\vec{0}, \vec{0}, \vec{0}}>$,
(c) 8 spinors of positive and 8 of negative chirality in the adjoint
representation of $E_8$ namely:
$|{\psi^+ \chi_8^+}>\otimes (|{\vec{0}, \vec{\alpha}, \vec{0}}>
	+ |{\vec{0}, \vec{0}, \vec{-\alpha}}>)$,
$|{\psi^- \chi^-_8}>\otimes (|{\vec{0}, \vec{\alpha}, \vec{0}}>
	- |{\vec{0}, \vec{0}, \vec{-\alpha}}>)$,
$|{\psi^- \chi_8^-}>\otimes X^{+R I}_{-1}|{\vec{0}, \vec{0}, \vec{0}}>$
and 
$|{\psi^+ \chi_8^+}>\otimes X^{-R I}_{-1}|{\vec{0}, \vec{0}, \vec{0}}>$.

The contributions of the spinors in the adjoint representation of $E_8$
cancel out in the anomaly computation as they come in pairs of 
opposite chirality. Since the spectrum in the twisted sector remains
unchanged, the theory is anomaly free at the 
enhanced symmetry point. At generic points the 
full spectrum matches with the ones for the 
compactification of $M$-Theory found earlier in this section. 
\sxn{Orientifolds of $K3\times K3\times \S^1$}
To obtain the two dimensional spectrum for this case, 
we now decompose the 
11-dimensional indices $M$ into $\mu$, $m$, $a$ and $x^{10}$, 
where $\mu$, $m$ and $a$  
denote the two dimensional space-time, and the two $K3$'s:
$K3^{(1)}$ and $K3^{(2)}$ respectively. Then the eleven-dimensional metric 
$G_{MN}$, upon compactification to two dimensions, gives among other
fields, a graviton $G_{\mu \nu}$, 58 scalars $G_{m n}$ appearing as the
moduli of the first $K3$, another 58 scalars appearing as the moduli
of the second $K3$ and one more scalar, namely $G_{{10}, {10}}$. The 
ten-dimensional three-form field $A_{M N P}$ gives, upon compactification,
22 scalars $A_{m n 10}$ and another 22 scalars $A_{a b 10}$ coming 
from the two-forms on $K3$'s. In the physical bosonic spectrum  
we have 160 scalars in two dimensions.

The number of fermions can once again be counted by using the index of the 
Dirac and Rarita-Schwinger operators on both the $K3$'s. 
The ten-dimensional
spinor $\Psi_{\bar{\mu}}^+$ now gives the following set of fields: 
(i) 4 positive chirality spin-$3\over 2$ fields
($\psi_{\mu}^+ \chi_{K_1}^- \chi_{K_2}^-$), 
(ii) Two sets of 80 spinors of negative chirality: ($\psi^- \chi_{K_1}^- 
\chi_{K_2 m}^+$ and $\psi^- \chi_{K_1 m}^+ \chi_{K_2}^-$). 
Similarly $\Psi_{\bar{\mu}}^-$ contributes in two 
dimensions, four spin-$3\over 2$ fields of negative chirality and 
two sets of 80 spin-$1\over 2$ fields of positive chiralities. The ten 
dimensional spin-$1\over 2$ fields $\Psi_{10}^+$ and $\Psi_{10}^-$
contributes, in two dimensions, 4 spin-$1\over 2$ fields of positive and
negative chiralities.

By cancelling the negative physical degrees of freedom of the gravitino
with the 4 spin-$1\over 2$ fields of positive and 4 of negative
chiralities, we are left with 160 spinors of positive and 160 of 
negative chiralities. By taking into account the scalars,
we have a $(4, 4)$ supersymmetric
model with its 40 scalar multiplets as the matter content. 
The theory is non-chiral and anomaly-free. 
The orientifold of $K3\times K3\times \S^1$ can be constructed in the 
same way as in the previous sections, with the $ \Z_2$ action on the 
two $K3$'s defined in an identical fashion. This action can be 
summarized as $A_{M N P}\rightarrow - A_{M N P}$,
$\psi_M \rightarrow \gamma^{11}\psi_M$ and  
$x^{10}\rightarrow - x^{10}$. Its action on individual $K3$'s 
is also described earlier. 

Using these results we find that there are 
96 scalar fields from the untwisted sector in this theory
in the physical bosonic spectrum. The 
$ \Z_2$ invariant fermion fields in two dimensions are as follows.
The ten-dimensional spinor $\Psi_{\bar{\mu}}^+$ gives the following
$ \Z_2$ invariant states in two dimensions (i) 4 spin-$3\over 2$
fields ($\psi_{\mu}^+ \chi_{k_1}^-\chi_{k_2}^-$) 
(ii) Spin-$1\over 2$ fields
$\psi^- \chi_{k_1}^-\chi_{k_2 m}^+$ and $\psi^- 
\chi_{k_1 m}^+\chi_{k_2}^-$.
$\Psi_{\bar{\mu}}^-$ contributes  $\psi^+ \chi_{k_1 m}^+\chi_{k_2}^-$
and  $\psi^+ \chi_{k_1}^-\chi_{k_2 m}^+$. These together give 
4 spin-$3\over 2$ fields of positive chirality, 96 spin-$1\over 2$ 
fields of negative chirality and and 64 spin-$1\over 2$ 
fields of positive chirality. $\Psi_{10}^-$ contributes only four 
negative chirality states ($\psi^- \chi_{k_1}^-\chi_{k_2}^-$) and 
$\Psi_{10}^+$ does not have any $ \Z_2$-invariant state. 
In this case we have 96
negative and 64 positive chirality spinors in two dimensions. 
Now we have a $(4, 0)$ supersymmetric theory in two
dimensions with  the spectrum,
\begin{itemize}
\item{$ (1 g_{\mu \nu}, 1 \phi, 4 \psi^+_\mu, 4 \psi^-$)}
\item{ $(96 \phi_L, 96\psi^-)$}
\item{ $(96 \phi_R, 64\psi^+)$}.
\end{itemize} 
The left (negative)-chirality modes in 
the untwisted sector are 
combined into 24 multiplets of the (4, 0) theory. The 
right (positive)-chirality
modes consist of 96 right-moving scalars and 64 
{\sf mw} spinors. The anomaly cancellation now requires 
128 {\sf mw} spinors of
positive chirality. It is noticed that this is exactly the number of 
the twisted sectors in this theory. Once again,
taking into account the fact that 
the individual twisted sector contribution remains the 
same as for $\T^9/ \Z_2$, we
get cancellation of anomaly in two dimensions.

We now discuss the enhanced symmetry point on 
$(K3\times K3\times \S^1)/ \Z_2$  orientifolds.
Once again the approach is to find out the states in the heterotic string
theory, identify the projections in these theories and obtain the 
invariant spectrum with respect to 
these projections. The $M$-Theory on 
$K3\times K3\times \S^1$ is equivalent to the heterotic string theory
on $K3\times \T^4$. As a result the states of the former can be 
re-expressed as the states of the latter one.
In this case, we obtain the two dimensional model 
by compactifying the heterotic string theory in two steps, first to 
six dimensions on $\T^4$ and then to two dimensions on $K3$. 
By defining coordinates
$X^{\tilde{\mu}}$, ($\tilde{\mu} = 0, \cdots ,5$), 
$X^i$ ($i \in \T^4$), $X^{R I}$ $ (I = 1, \cdots ,16)$ as the
bosonic worldsheet coordinates and 
$\psi^{L \tilde{\mu}}$ and $\psi^{L i}$ as the fermionic coordinates 
on the worldsheet we get the 
following set of states in this theory in six dimensions: 
(i) Graviton  ($G_{\tilde{\mu}\tilde{\nu}}$), antisymmetric 
tensor ($B_{\tilde{\mu}\tilde{\nu}}$) and the dilaton $\phi$ arising
as various combinations of the states:
$\psi^{L \tilde{\mu}}_{-{1\over 2}}|{0}>\otimes 
X^{R \tilde{\nu}}_{-1}|\vec{0}, \vec{0}, \vec{0}>$,
(ii) 16 scalars $\phi^{i j}$ which are also the singlets of 
$E_8\times E_8$:
$\psi^{L i}_{-{1\over 2}}|{0}>\otimes 
X^{R j}_{-1}|\vec{0}, \vec{0}, \vec{0}>$,
(iii) 4 scalars in the adjoint representation of $E_8\times E_8$ 
written as:  
(a) $\psi^{L i}_{-{1\over 2}}|{0}>\otimes 
X^{R I}_{-1}|\vec{0}, \vec{0}, \vec{0}>$,
(b) $\psi^{L i}_{-{1\over 2}}|{0}>\otimes 
|{\vec{0}, \vec{\alpha}, \vec{0}}>$, and 
(c) $\psi^{L i}_{-{1\over 2}}|{0}>\otimes 
|{\vec{0}, \vec{0}, \vec{\alpha}}>$,

Upon further compactification to two dimensions on $K3$ we obtain from
$G_{\tilde{\mu} \tilde{\nu}}$, a graviton $G_{\mu \nu}$ and  
58 scalars $G_{m n}$.
$B_{\tilde{\mu} \tilde{\nu}}$ in six dimensions 
gives 22 scalars $B_{m n}$. In addition we also have the 
six dimensional scalars.
Counting of the states leads to the bosonic spectrum
of $M$-Theory at the enhanced 
symmetry point. This consists of the graviton, 97 scalars as singlets
of $E_8\times E_8$ and 8 set of scalars in the adjoint of 
$E_8\times E_8$. At a generic point in the moduli space we have  
161 scalars which matches with the ones presented earlier in this 
section for the $M$-Theory compactification. 

The action of $ \Z_2$ on the ten dimensional field as well as on $\T^4$ 
is defined exactly in the same manner as in the previous sections.
Its action on $K3$ is the same as the one mentioned earlier. 
The bosonic
states of the heterotic string theory surviving the projection
are the graviton $G_{\mu \nu}$, $\phi$,
34 scalars $G_{m n}$, 14 scalars
$B_{m n}$, 16 scalars $\phi^{i j}$, all appearing as singlets of 
$E_8$. In addition we have 4 copies of scalars in 
the adjoint of $E_8$ appearing as the set of states:
(i) $\psi^{L i}_{-{1\over 2}}|{0}>\otimes 
X^{+R I}_{-1}|\vec{0}, \vec{0}, \vec{0}>$ and
(ii) $\psi^{L i}_{-{1\over 2}}|{0}>\otimes 
(|\vec{0}, \vec{\alpha}, \vec{0}>- 
|\vec{0}, \vec{0}, \vec{-\alpha}>)$.

The fermionic spectrum is found by writing down the 
fermionic states in the heterotic language: 
(i) Gravitinos and the dilatinos in six dimensions appearing
as the decomposition of the state:  
$|\bar{\psi}>\otimes X^{R \tilde{\nu}}_{-1}
 |{\vec{0}, \vec{0}, \vec{0}}>$
into six and four dimensional ($\T^4$) parts. 
(ii) $E_8\times E_8$ singlet spinors:
$|{\bar{\psi}}>\otimes X^{R i}_{-1}|{\vec{0}, \vec{0}, \vec{0}}>$
and (iii) spinors in the adjoint of $E_8\times E_8$:
(a) $|{\bar{\psi}}>\otimes X^{R I}_{-1}|{\vec{0}, \vec{0}, \vec{0}}>,$
(b) $|{\bar{\psi}}>\otimes |{\vec{0}, \vec{\alpha}, \vec{0}}>$ and
(c) $|{\bar{\psi}}>\otimes |{\vec{0}, \vec{0}, \vec{\alpha}}>$. 
Further compactification of these spinors to two dimensions is done in
the same way as in sections-(2) and (3). For example, a state 
$|\bar{\psi}>\otimes X^{R \tilde{\nu}}_{-1}|\vec{0}, \vec{0}, \vec{0}> 
\equiv |{\bar{\psi}}_{\tilde{\mu}}^+>$ is decomposed as a
spin-$3\over 2$ states $|\psi_{\mu}^-\chi_k^-\chi_4^+>$,
$|\psi_{\mu}^+\chi_k^-\chi_4^->$, and spin-$1\over 2$ states
$|\psi^+\chi_{k m}^+\chi_4^+>$, $|\psi^-\chi_{k m}^+\chi_4^->$.
By using the Dirac and Rarita-Schwinger 
index on $K3$ we obtain 4 spin-$3\over 2$ {\sf mw} fermions: 
($|\psi_{\mu}^+\chi_k^- \chi_4^->$) which are even under $ \Z_2$. 
They also give us 
32 positive and 48 negative chirality $ \Z_2$-even spin-$1\over 2$ 
states and same number of odd ones. 
For other ten dimensional states,
the decomposition of the $E_8\times E_8$ singlet states (ii) give
16 even states of negative chirality and 16 odd ones of 
positive chirality. For the spin-$1\over 2$ 
$ \Z_2$-even states which are in the 
adjoint of the $E_8\times E_8$ gauge group, we have 4 of them 
of positive and 4 of negative chirality. 
By counting all the even states we get 4 gravitinos appearing as
positive chirality {\sf mw} spinors. Among the spin-$1\over 2$ 
negative chirality states, we
have 64 singlets of $E_8$ and 4 adjoints. Among the positive
chirality ones are the 32 singlets and 4 adjoints of $E_8$. Since the
adjoint representation appears as a pair of opposite chirality states,
anomaly cancellation is maintained. 

\sxn{Conclusions}

To conclude, we have presented $M$-Theory compactifications to two 
dimensions. We have shown that compactification on orientifolds leads
to chiral theories in two dimensions. By taking into account the 
twisted sector states, anomaly cancellation is shown in many 
examples. The index theorem on $K3$ and a $ \Z_2$ 
abelian automorphism of these manifolds 
play an important role in the computations. The anomaly 
cancellation at some enhanced symmetry points have also been shown. 
Our results confirm the number of twisted 
sectors and their contributions as given in \cite{mukhi,wit2}. 
We have also shown 
that the states obtained are always in 
the representations of the unbroken supersymmetry
algebra. 

In this paper we have been able to obtain (16,0), (8, 0) and (4, 0)
models through the orientifold compactifications. It will be  
interesting to extend this analysis to obtain (2, 0) and (1,0) chiral
models in two dimensions. Considering the Calabi-Yau space for this purpose
does not seem to change the number of supersymmetries, as they have to 
be combined with the extra $\T^3$ leaving once again $1\over 4$ supersymmetry
unbroken. Other compactifications such as on Joyce manifolds may have to
be considered for this purpose. Other option may be to find other 
automorphisms of $K3$ which break the supersymmetries further. 
It may also be interesting to analyze 
the moduli space of these theories and find out the Geroch group 
structure\cite{biswas}. Normally the symmetries of the three 
dimensional theories are affinized on compactification 
to two dimensions. One, however, has to 
write down the effective action for the purpose of examining 
the Geroch symmetry structure and therefore 
it may be useful to understand these models in terms of the 
type IIA and type IIB string compactifications\cite{tsen}. 

\end{document}